**Pedestrian Safety and Traffic Operations Around Near-Side Versus Far-Side Transit Stops: Emerging Observational Evidence from Utah**

**Fariba Soltani Mandolakani**
Department of Civil and Environmental Engineering
Utah State University, Logan, UT 84322-4110
Email: fariba.soltani@usu.edu
ORCID: 0009-0005-9089-0514

**Atul Subedi**
Department of Civil and Environmental Engineering
Utah State University, Logan, UT, 84322-4110
Email: atul.subedi@usu.edu
ORCID: 0000-0002-0398-0599

**Patrick A. Singleton** (corresponding author)
Department of Civil and Environmental Engineering
Utah State University, Logan, UT, 84322-4110
Email: patrick.singleton@usu.edu
ORCID: 0000-0002-9319-2333

**Michelle Mekker**
Department of Civil and Environmental Engineering
Utah State University, Logan, UT, 84322-4110
Email: mekker@highstreetconsulting.com

**ABSTRACT**

This research project's objective was to investigate the impacts of transit stop location (near-side versus far-side) on pedestrian safety and traffic operations. Three different video-based behavioral observation data collections at signalized intersections in Utah were utilized, studying: (1) transit vehicle stop events and transit rider crossing behaviors and vehicle conflicts; (2) pedestrian conflicts with right-turning vehicles (driver/pedestrian reactions, conflict severity); and (3) pedestrian crossing behaviors (crossing location, crossing behaviors). These outcomes were statistically compared for near-side versus far-side transit stop locations. Far-side transit stops appear better for general traffic operations. Although transit departure delays are more likely and impactful at far-side stops, actions can be taken to improve transit operations there. On the other hand, far-side transit stops appear to be worse for pedestrian safety, corroborating prior crash-based research findings. Specifically, conflicts at far-side stops were more severe, and drivers were less likely to slow/stop for pedestrians. Reconciling these differing findings likely requires improving pedestrian safety at some far-side transit stops, and prioritizing safety over operational efficiency at other near-side transit stops.

## INTRODUCTION

Pedestrian injuries and fatalities are rising in the United States overall, and in most states specifically. The Governors Highway Safety Association (*1*) reports a 75% increase in pedestrian fatalities since 2010, now accounting for 18% of all traffic deaths. Most fatalities occur on non-freeway arterials, particularly at intersections. In 2022, the state of Utah reported 52 fatal and 819 injury crashes involving pedestrians (*2*). Pedestrians, as vulnerable road users, are more likely to be injured or killed in collisions.

Research in the US and in other countries has found that intersections with more nearby transit stops tend to experience more pedestrian crashes. In the U.S., this trend has been observed in California (*3*), Florida (*4*), Maryland (*5*), New York (*6*), North Carolina (*7,8*), Texas (*9*), and Washington (*10*). Internationally, similar findings were reported in Canada (*11,12*) and Peru (*13*). These relationships persist even when controlling for pedestrian exposure, a strong predictor of crash frequencies (*14*).

Mirroring these findings, research in Utah has also shown more pedestrian crashes at intersections with transit stops. Recent studies for the Utah Department of Transportation (UDOT) found higher crash rates at intersections with nearby transit stops (*15–17*), especially far-side stops. Specifically, at least one recent study in Utah (*14*) found that far-side stops increase pedestrian crashes more than near-side stops. It is uncertain why far-side transit stops might be less safe for pedestrians, especially because Utah transit agencies consider near-side stops more problematic due to blocked views and increased conflicts. Other studies suggest that far-side stops may improve safety by encouraging pedestrians to cross behind the transit vehicle (*18,19*). Far-side stops are also operationally advantageous, reducing vehicle queuing and delays, and are preferred at complex intersections (*20–24*). Yet, overall, research on the safety implications of transit stop placement (near-side vs. far-side) is limited. This project aims to understand the impact of transit stop locations on pedestrian safety.

Pedestrian behaviors contribute to around 50% of crashes, including improper crossing, darting, and inattentiveness (*25*). These behaviors are critical at intersection transit stops but lack comprehensive documentation. Previous UDOT research examined pedestrian behaviors (*26,27*) but did not focus on transit stop locations. Studying pedestrian behaviors and interactions near transit stops could help to inform safety improvements and operational adjustments. Past studies have also not fully considered or balanced the safety and operational impacts of transit stop placement at intersections. Near-side stops promote crosswalk use but may increase delays, while far-side stops might reduce operational efficiency. This project seeks to understand both safety and operational implications of near-side versus far-side transit stops.

The primary objective of this research is to investigate the impacts of transit stop location (near-side vs. far-side of intersections) on both pedestrian safety and traffic operations. The research findings may offer recommendations for improving both pedestrian safety and traffic operations at intersections with transit stops.

The paper is organized as follows. The "Data and Methods" section describes the study area in Utah, the data collections utilized, and the analysis methods applied. The following section presents the results, including insights from interactions between transit vehicles, pedestrians, and turning vehicles. A conclusion "Discussion" section provides key findings related to the research objective, offers recommendations, and notes study limitations and opportunities for future work.

## DATA AND METHODS

**Study Area and Locations**

The data for this study were collected at signalized intersections with nearby transit stops (within 300 ft) in the state of Utah. A medium-sized state located in the Intermountain West, Utah was the 31st-most populous state in 2020, but it had the fastest percentage growth (18%) of any US state between 2010 and 2020 (*28*). Around 75% of the population lives in four counties along what is called the Wasatch Front region. Most of the study locations were along this urbanized Wasatch Front corridor, from Ogden through Salt Lake City to Provo, although several study locations were in other cities throughout the state (near or in Logan, St. George, Cedar City, and Moab).

Figure 1 shows the study locations within the context of the state of Utah (left) and the Wasatch Front region (right). As described in the following sections, this study utilized three different data collection efforts: data collection 1 followed transit passengers before/after boarding; data collection 2 studied pedestrians involved in conflicts with right-turning vehicles; and data collection 3 observed all pedestrians crossing the street. The study locations associated with each data collection effort are shown in different colors on the maps of Figure 1.

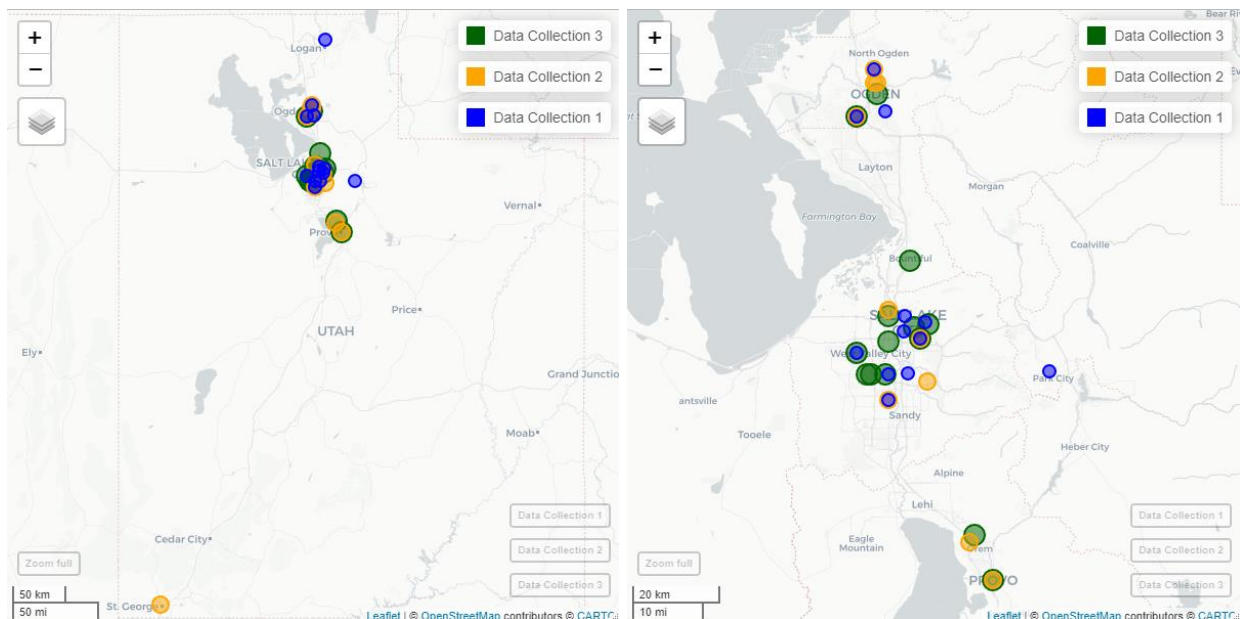

**Figure 1: Maps of Utah (left) and the Wasatch Front (right), showing study locations**

**Data Collection 1**

Original data were collected to identify the impacts of near-side versus far-side transit stop locations on pedestrian safety and traffic operations at intersections. Potential study locations were selected from signalized intersections in Utah with nearby transit stops, utilizing data from various transit agencies—Utah Transit Authority (UTA), the Cache Valley Transit District (CVTD), and Park City Transit—to create a spatial database of transit stops, their average daily trip frequency, and (when available) their average daily ridership. These transit data were merged with traffic signal information to create a spatial database that included the number of near-side, far-side, and total transit stops (along with average daily trip frequency and ridership at each location). A 300 ft buffer from the center of each intersection was used to identify if transit stop was "near" the intersection, with near-side stops on the "inbound" side and far-side stops on the "outbound" side. Specific study locations were chosen based on the presence of existing overhead traffic cameras (operated by UDOT) nearby, sufficient ridership or trip frequency, and expert recommendations from a research project advisory committee (including transit agency staff members). Further, an effort was made to balance sites regionally and include diverse operational contexts, resulting in 15 transit stops at 13 intersections being selected for data collection. At each study location, around two days of video were recorded using UDOT-operated overhead traffic cameras. Most videos were recorded on weekdays in June 2023, but some locations used previously recorded videos from August 2019, June 2020, November 2021, and late May/early June 2022. Once the videos were recorded, data extraction was done by trained observers, who recorded details about each transit stop event, as well as each passenger boarding or alighting and (if visible from the video) information about crossing behaviors and interactions/conflicts while crossing. A detailed survey form and training document were created to ensure consistent data collection.

Videos recorded at 13 intersections (6 with a near-side transit stop, 6 with a far-side transit stop, and 1 with both a far-side and near-side stop) resulted in 1,592 observations of transit stop events and 2,582 observations of pedestrian transit riders, of which 1,115 had an observed crossing and 60 had an observed conflict with a motor vehicle. Table 1 reports descriptive statistics for the variables contained within the dataset that was originally collected for this study. More information about data collection 1 is available from the authors upon request.

**Table 1: Descriptive statistics for data collection 1**

| Variable | # | % | Mean | SD |
|---|---|---|---|---|
| **Transit stop event information ($N = 1{,}592$)** | | | | |
| Transit stop location: Far-side | 1,249 | 78.45 | | |
| Near-side | 343 | 21.55 | | |
| Number of people: Boarded (got on to) the transit vehicle | | | 0.68 | 1.03 |
| Alighted (got off of) the transit vehicle | | | 0.74 | 1.10 |
| The transit vehicle was delayed by traffic when arriving at the stop | 22 | 1.38 | | |
| The transit vehicle was delayed by traffic when leaving the stop | 227 | 14.26 | | |
| Other traffic was delayed by the transit vehicle while it was stopped | 82 | 5.15 | | |
| Some other vehicles changed lanes in order to pass the stopped transit vehicle | 138 | 8.67 | | |
| The transit vehicle stopped to pick-up/drop-off passengers, but not at or near the stop location | 22 | 1.38 | | |
| The transit vehicle was blocking a driveway or intersection while it was stopped | 3 | 0.19 | | |
| Number of vehicles: Between the transit vehicle and the stop location, while it was delayed by traffic when arriving at the stop | | | 0.03 | 0.30 |
| Delayed by the transit vehicle, while it was stopped | | | 0.10 | 0.57 |
| Changed lanes in order to pass the stopped transit vehicle, while it was stopped | | | 0.17 | 0.68 |
| Passed the transit vehicle, while it was delayed by traffic when leaving the stop | | | 0.33 | 1.06 |
| Dwell time (sec) (vehicle departure time − vehicle arrival time) | | | 32.69 | 34.64 |
| **Pedestrian information ($N = 2{,}582$)** | | | | |
| Transit stop location: Far-side | 2,110 | 81.72 | | |
| Near-side | 472 | 18.28 | | |
| Pedestrian action: Alighting | 1,525 | 59.06 | | |
| Boarding | 1,057 | 40.94 | | |
| Age: Adult of unknown age | 996 | 38.59 | | |
| Child | 31 | 1.20 | | |
| Teenager | 326 | 12.63 | | |
| Young adult | 1,010 | 39.13 | | |
| Middle-aged adult | 175 | 6.78 | | |
| Older adult | 43 | 1.67 | | |
| Gender: Unknown gender | 2,366 | 91.63 | | |

| | | |
|---|---:|---:|
| Female | 69 | 2.67 |
| Male | 147 | 5.69 |
| Other characteristics: Traveling with 1+ other person(s) | 401 | 15.53 |
|    Carrying load (larger than small purse or backpack) | 58 | 2.25 |
|    Stroller | 13 | 0.50 |
|    Wheelchair | 14 | 0.54 |
|    Skateboard | 11 | 0.43 |
|    Scooter | 4 | 0.15 |
|    Bicycle | 11 | 0.43 |
|    Distracted (by phone, headphones, conversation, etc.) | 39 | 1.51 |
|    Other | 441 | 17.08 |
| Alighting person ($N = 1,525$): Crossed at an intersection or a marked crossing | 684 | 44.85 |
|    Crossed mid-block, away from an intersection or a marked crossing | 19 | 1.25 |
|    Did not cross a street; turned a corner instead | 25 | 1.64 |
|    Did not cross a street; walked away from the transit stop along the street | 190 | 12.46 |
|    Did not cross a street; went to an adjacent land use (business, home, etc.) | 128 | 8.39 |
|    Did not leave; stayed at the transit stop to board another transit vehicle | 15 | 0.98 |
|    Cannot see from the view of the video | 396 | 25.97 |
|    Other | 68 | 4.46 |
| Boarding person ($N = 1,057$): Crossed at an intersection or a marked crossing | 395 | 37.37 |
|    Crossed mid-block, away from an intersection or a marked crossing | 17 | 1.61 |
|    Did not cross a street; turned a corner instead | 22 | 2.08 |
|    Did not cross a street; walked towards the transit stop along the street | 151 | 14.29 |
|    Did not cross a street; came from an adjacent land use (business, home, etc.) | 255 | 24.12 |
|    Did not arrive; was already at the transit stop after alighting another transit vehicle | 45 | 4.26 |
|    Cannot see from the view of the video | 149 | 14.10 |
|    Other | 23 | 2.18 |
| **Pedestrian crossing information ($N = 1,115$)** | | |
| Transit stop location: Far-side | 907 | 81.35 |
|    Near-side | 208 | 18.65 |
| Street crossed: The main street | 896 | 80.72 |
|    A side street | 213 | 19.19 |
|    Other | 1 | 0.09 |
| Pedestrian behavior: Was outside of the crosswalk markings for most if not all of the crossing | 45 | 4.04 |
|    Changed speed (e.g., walk to run, or run to walk) | 25 | 2.24 |
|    Paused in the middle of the street | 9 | 0.81 |
|    Seemed distracted by phone or something else | 8 | 0.72 |
|    Crossed just in front of the transit vehicle (if present) | 5 | 0.45 |
|    Crossed just behind the transit vehicle (if present) | 10 | 0.90 |
| **Pedestrian–vehicle conflict information ($N = 60$)** | | |
| Transit stop location: Far-side | 37 | 61.67 |
|    Near-side | 23 | 38.33 |
| Conflict location: When the person was entering (or starting to cross) the street | 28 | 46.67 |
|    When the person was in the middle of the street | 22 | 36.67 |
|    When the person was exiting (or finishing crossing) the street | 10 | 16.67 |
|    Other | 0 | 0.00 |
| Vehicle movement: Driving straight | 32 | 55.17 |
|    Turning right | 17 | 29.31 |
|    Turning left | 9 | 15.52 |
|    Other | 0 | 0.00 |
| Driver reaction: No obvious reaction | 38 | 63.33 |
|    Driver fully stopped | 10 | 16.67 |
|    Driver slowed down | 11 | 18.33 |
|    Driver sped up | 0 | 0.00 |

| | # | % | Mean | SD |
|---|---|---|---|---|
| Driver swerved | 0 | 0.00 | | |
| Cannot see from the view of the video | 1 | 1.67 | | |
| Pedestrian reaction: No obvious reaction | 26 | 43.33 | | |
| Stopped and waited for the vehicle | 27 | 45.00 | | |
| Slowed down to avoid collision | 0 | 0.00 | | |
| Sped up or ran to avoid collision | 3 | 5.00 | | |
| Changed direction | 1 | 1.67 | | |
| Cannot see from the view of the video | 3 | 5.00 | | |
| Encroachment time (sec): abs (vehicle time − pedestrian time at the conflict point) | | | 1.89 | 0.93 |

**Data Collection 2**

Data collection 2 was originally obtained for a different project focused on pedestrian safety (*27*), studying observed conflicts between pedestrians (not necessarily transit riders) and right-turning motor vehicles occurring at signalized intersection corners. For the purposes of this study, the data were filtered to include only study locations with either a near-side or a far-side transit stop on the corner that was observed. This was done in an attempt to provide as much consistency as possible between the datasets: both datasets include pedestrians (in some or all cases) interacting with vehicles at intersections with near-side and/or far-side transit stops. Nevertheless, some key differences remain: The right-turn dataset (data collection 2) included all pedestrians interacting with right-turning vehicles at corners with transit stops. In contrast, data collection 1 included only transit riders (not all pedestrians) but did count pedestrians crossing the street even if they did not interact with a right-turning vehicle.

After filtering the right-turn dataset, information about 807 conflicts between pedestrians and right-turning vehicles at 11 signalized intersection corners in Utah (3 with a near-side transit stop, 7 with a far-side transit stop, and 1 with both a far-side and near-side stop) remained. For the purposes of this study, several pedestrian safety related outcomes were considered—such as crossing location, conflict severity, and both pedestrian and driver reactions to conflicts—as detailed in Table 2.

**Table 2: Descriptive statistics for data collection 2**

| *Variable* | # | % | *Mean* | *SD* |
|---|---|---|---|---|
| **Pedestrian–right-turning vehicle conflict information (*N* = 807)** | | | | |
| Transit stop location: Far-side | 661 | 81.91 | | |
| Near-side | 146 | 18.09 | | |
| Group size (# people) | | | 1.20 | 0.62 |
| Age: Child | 21 | 2.60 | | |
| Teenager | 68 | 8.43 | | |
| Young adult | 323 | 40.02 | | |
| Middle-aged adult | 328 | 40.64 | | |
| Older adult | 24 | 2.97 | | |
| Adult of unknown age | 84 | 10.41 | | |
| Gender: Male | 537 | 66.54 | | |
| Female | 234 | 29.00 | | |
| Unknown gender | 89 | 11.03 | | |
| Other characteristics: Carrying load | 33 | 4.09 | | |
| Stroller | 7 | 0.87 | | |
| Wheelchair | 4 | 0.50 | | |
| Skateboard | 20 | 2.48 | | |
| Scooter | 21 | 2.60 | | |
| Bicycle | 135 | 16.73 | | |
| Distracted (phone, headphones, conversation, etc.) | 9 | 1.12 | | |
| Crosswalk: First crosswalk | 122 | 15.12 | | |
| Second crosswalk | 685 | 84.88 | | |
| Crossing location: In the crosswalk or the crosswalk area | 790 | 97.89 | | |
|    Mid-block, away from the crosswalk | 13 | 1.61 | | |
|    In the middle of the intersection | 4 | 0.50 | | |
| Crossing direction: Leaving curb | 536 | 66.42 | | |
|    Approaching curb | 271 | 33.58 | | |
| Pedestrian reaction: No obvious reaction | 725 | 89.84 | | |
|    Stopped and waited for the vehicle | 36 | 4.46 | | |
|    Slowed down to avoid collision | 8 | 0.99 | | |
|    Sped up to avoid collision | 18 | 2.23 | | |
|    Ran to avoid collision | 10 | 1.24 | | |
|    Changed direction | 10 | 1.24 | | |
| Right-turn queue length (# vehicles) | | | 1.91 | 1.69 |
| Stopping location: Did not stop | 489 | 60.59 | | |
|    Before the first crosswalk | 112 | 13.88 | | |
|    Inside the first crosswalk | 145 | 17.97 | | |
|    Between the first and second crosswalks | 61 | 7.56 | | |
|    Inside the second crosswalk | 0 | 0.00 | | |
| Driver reaction: No obvious reaction | 415 | 51.43 | | |
|    Driver fully stopped | 167 | 20.69 | | |
|    Driver slowed down | 208 | 25.77 | | |
|    Driver sped up | 15 | 1.86 | | |
|    Driver swerved | 2 | 0.25 | | |
| Vehicle type: Small (sedan, motorcycle) | 333 | 41.26 | | |
|    Medium (SUV, pickup truck, van) | 451 | 55.89 | | |
|    Large (large truck, vehicle pulling a trailer, bus) | 23 | 2.85 | | |
| Encroachment time (sec): abs(vehicle time − pedestrian time at the conflict point) | | | 5.58 | 2.30 |
| Conflict severity: Low (5-10 sec) | 409 | 50.68 | | |
|    Mild (4-5 sec) | 242 | 29.99 | | |
|    High (0-3 sec) | 156 | 19.33 | | |

**Data Collection 3**

     Similarly, data collection 3 was originally obtained for a different project focused on pedestrian safety (*26*), specifically studying pedestrian crossing behaviors at signalized intersections. For the purposes of this study, the data were filtered to include only study locations with a near-side and/or a far-side transit stop on one of the two corners connecting to the crosswalk that was observed. This was done in an attempt to provide as much consistency as possible with data collection 1: both datasets include pedestrians (in some or all cases) crossing at intersections from or to corners at with near-side and/or far-side transit stops. Nevertheless, some key differences remain: The pedestrian crossing behavior dataset (data collection 3) included all pedestrians crossing to/from at corners with transit stops. In contrast, data collection 1 included only transit riders (not all pedestrians).

     After filtering the pedestrian crossing behavior dataset, information about 2,611 pedestrian crossing events at 14 signalized intersections in Utah (2 connecting corners with a near-side transit stop, 11 with a far-side transit stop, and 1 with both a far-side and near-side stop) remained. For the purposes of this study, several pedestrian safety outcomes were considered—such as crossing location, crosswalk marking adherence, crossing behaviors, and crossing obstacles—as detailed in Table 3.

**Table 3: Descriptive statistics for data collection 3**

| Variable | # | % | Mean | SD |
|---|---|---|---|---|
| **Pedestrian crossing information (*N* = 2,611)** | | | | |
| Transit stop location: Far-side | 2,172 | 83.19 | | |
|    Near-side | 439 | 16.81 | | |
| Group size (# people) | | | 1.24 | 0.66 |
| Age: Child | 40 | 1.53 | | |
|    Teenager | 106 | 4.06 | | |
|    Young adult | 951 | 36.42 | | |
|    Middle-aged adult | 941 | 36.04 | | |
|    Older adult | 30 | 1.15 | | |
|    Adult of unknown age | 630 | 24.13 | | |
| Gender: Male | 1,563 | 59.86 | | |
|    Female | 674 | 25.81 | | |
|    Unknown gender | 605 | 23.17 | | |
| Other characteristics: Carrying load | 29 | 1.11 | | |
|    Stroller | 16 | 0.61 | | |
|    Wheelchair | 19 | 0.73 | | |
|    Skateboard | 36 | 1.38 | | |
|    Scooter | 51 | 1.95 | | |
|    Bicycle | 359 | 13.75 | | |
|    Other | 77 | 2.95 | | |
| # other people waiting | | | 0.17 | 0.55 |
| # vehicles passing (past 10 sec) | | | 4.12 | 4.65 |
| # vehicles passing (next 10 sec) | | | 3.57 | 4.28 |
| Waiting behaviors: Pressed pedestrian push-button | 1,380 | 52.85 | | |
|    Paced or otherwise seemed impatient | 135 | 5.17 | | |
|    Left waiting area without crossing street | 39 | 1.49 | | |
|    Other | 72 | 2.76 | | |
| Waiting time (sec) | | | 30.04 | 31.25 |
| Crossing location: In the crosswalk or the crosswalk area | 2,488 | 95.29 | | |
|    Mid-block, away from the crosswalk | 82 | 3.14 | | |
|    In the middle of the intersection | 2 | 0.08 | | |
|    Other | 39 | 1.49 | | |
| # other people crossing (same direction) | | | 0.24 | 0.63 |
| # other people crossing (opposite direction) | | | 0.20 | 0.55 |
| Crosswalk markings: Stayed within the crosswalk markings for all or almost the whole crossing | 2,216 | 84.87 | | |
|    Stepped outside of the crosswalk markings for part of the crossing | 220 | 8.43 | | |
|    Was outside of the crosswalk markings for most if not all of the crossing | 165 | 6.32 | | |
|    Other | 2 | 0.08 | | |
| Crossing behaviors: Changed speed (e.g., walk to run, or run to walk) | 153 | 5.86 | | |
|    Paused in the middle of the street | 59 | 2.26 | | |
|    Seemed distracted by phone or something else | 10 | 0.38 | | |
|    Other | 50 | 1.91 | | |
| Crossing obstacles: Car blocking the crosswalk | 121 | 4.63 | | |
|    Snow pile, water puddle, or debris | 0 | 0.00 | | |
|    Other | 4 | 0.15 | | |
| Crossing time (sec) | | | 17.07 | 7.21 |

**Analysis Methods**

     The study's objective was to investigate the impact of transit stop location (near-side versus far-side) on pedestrian safety and traffic operations. This involved comparing various dependent

variables related to traffic operations and pedestrian safety for both near-side and far-side transit stops using statistical tests.

For continuous dependent variables (measured as integers or real numbers), Welch's t-test was used. This test compares the mean values of two populations (near-side and far-side) and is suitable when the samples have unequal variances and sizes.

$$t = \frac{\bar{X}_1 - \bar{X}_2}{\sqrt{s_{\bar{X}_1}^2 + s_{\bar{X}_2}^2}} \text{ and } s_{\bar{X}_i} = \frac{s_i}{\sqrt{N_i}} \text{ and } v = \left(\frac{s_1^2}{N_1} + \frac{s_2^2}{N_2}\right)^2 \bigg/ \left(\frac{s_1^4}{N_1^2 v_1} + \frac{s_2^4}{N_2^2 v_2}\right)$$

where $\bar{X}_i$ is the sample mean, $s_i$ is the sample standard deviation, $N_i$ is the sample size, and $v_i = N_i - 1$ is the degrees of freedom for group $i$. Test statistic $t$ is then compared against the t-distribution with degrees of freedom $v$ to determine the statistical significance of the mean difference ($\bar{X}_1 - \bar{X}_2$).

For binary dependent variables (true/false categories), Fisher's exact test was used. Like a chi-square test, this test compares the proportions of one category in two populations (near-side and far-side) and is appropriate for small sample sizes. It produces a test statistic (odds ratio) and a p-value, indicating the probability that the observed ratio of proportions is equal in the population. This determines the statistical significance of the difference in proportions.

**RESULTS**

Results of the data analyses of key outcomes related to traffic operations and pedestrian safety, compared for near-side and far-side transit stops, are presented and described in the following sections. The sections are organized according to each of the three data collections.

**Data Collection 1**

*Traffic Operations*

Three types of dependent variables related to traffic operations were analyzed: (a) Whether the *transit vehicle was delayed* by other traffic, and by how much, measured in number of vehicles; (b) Whether *other traffic was delayed* by the transit vehicle; and by how much, measured in number of vehicles; and (c) Whether the *stopping location* of the transit vehicle (away from the stop location, or blocking a driveway or intersection) implied potential adverse operational impacts. Results are shown in Table 4.

There were statistically significant differences in other *traffic's impacts to transit operations*. Transit vehicles experienced more frequent and severe arrival delays due to other traffic at near-side stops (2.9% occurrence, 0.09 vehicles on average) compared to far-side stops (1.0%, 0.01 vehicles). Departure delays (excluding signal-induced delays) were significantly higher at far-side stops (17.6%, 0.42 vehicles) than at near-side stops (2.0%, 0.02 vehicles). It was more likely for the transit vehicle to stop to pick-up/drop-off passengers but not at the actual stop location for near-side stops (3.2%) than for far-side stops (0.9%). To summarize, for transit vehicles, arrival delays and impacts were more likely and more severe at near-side stops, but departure delays were much more likely and impactful at far-side stops.

There were also some statistically significant differences in *transit vehicles' impacts on other traffic* for near-side versus far-side stops. Transit vehicles delayed other traffic more often and more substantially at near-side stops (14.0% occurrence, 0.34 vehicles on average) than at far-side stops (2.7%, 0.04 vehicles). Other vehicles changed lanes to bypass transit vehicles similarly often at both stop types (8.7–8.8%), though slightly more vehicles were involved at far-side stops (0.19) versus near-side stops (0.12). It was more common for transit vehicles to block a driveway or intersection at near-side stops (0.6%) than at far-side stops (0.1%), but this difference was not statistically significant. To summarize, for other traffic, transit vehicles delayed other vehicles much more often and more significantly at near-side stops, whereas somewhat more passing was observed at far-side stops.

**Table 4: Results for data collection 1, traffic operations**

| Traffic operations | Far-Side | Near-Side | Far-Side | Near-Side | Statistic | df | p-value |
|---|---|---|---|---|---|---|---|
| Dependent variable | Sample size | | Proportion (true) | | Fisher's exact test | | |
| The transit vehicle was delayed by traffic when arriving at the stop | 1,249 | 343 | 0.0096 | 0.0292 | 3.09 | -- | 0.015 |
| The transit vehicle was delayed by traffic when leaving the stop | 1,249 | 343 | 0.1761 | 0.0204 | 0.10 | -- | 0.000 |
| Other traffic was delayed by the transit vehicle while it was stopped | 1,249 | 343 | 0.0272 | 0.1399 | 5.81 | -- | 0.000 |
| Some other vehicles changed lanes in order to pass the stopped transit vehicle | 1,249 | 343 | 0.0865 | 0.0875 | 1.01 | -- | 0.914 |
| The transit vehicle stopped to pick-up/drop-off passengers, but not at or near the stop location | 1,249 | 343 | 0.0088 | 0.0321 | 3.73 | -- | 0.003 |
| The transit vehicle was blocking a driveway or intersection while it was stopped | 1,249 | 343 | 0.0008 | 0.0058 | 7.31 | -- | 0.119 |
| Dependent variable | Sample size | | Mean (#) | | Welch's t-test | | |
| Number of vehicles between the transit vehicle and the stop location, while it was delayed by traffic when arriving at the stop | 1,249 | 343 | 0.0128 | 0.0875 | -2.48 | 363 | 0.014 |
| Number of vehicles passed the transit vehicle, while it was delayed by traffic when leaving the stop | 1,249 | 343 | 0.4171 | 0.0233 | 11.27 | 1440 | 0.000 |
| Number of vehicles delayed by the transit vehicle, while it was stopped | 1,249 | 343 | 0.0392 | 0.3411 | -5.06 | 352 | 0.000 |
| Number of vehicles changed lanes in order to pass the stopped transit vehicle, while it was stopped | 1,249 | 343 | 0.1857 | 0.1224 | 1.98 | 897 | 0.048 |

*Pedestrian Safety: Pedestrian Crossing Behaviors*

Two types of pedestrian safety dependent variables related to crossing behaviors were examined: (a) Whether the *pedestrian crossing location* was at an intersection or a marked crossing, or if it was mid-block (i.e., not at an intersection or a marked crossing); and (b) Multiple *pedestrian crossing behaviors*, observed while crossing, such as: was outside of the crosswalk markings, changed speed, paused in the middle of the street, seemed distracted, and crossed just in front of or behind the transit vehicle (if present). Results are shown in Table 5.

For *pedestrian crossing locations*, slightly more mid-block pedestrian crossings were observed at far-side stops (3.4%) than at near-side stops (2.4%), although the difference was not statistically significant. Similarly, the small differences observed for most *pedestrian crossing behaviors* were not statistically significant: crossing outside crosswalk markings, pausing, and appearing distracted were slightly more common at near-side stops; while changing speed and crossing just behind the transit vehicle were slightly more common at far-side stops. The only significant difference was the higher share of pedestrians crossing directly in front of the transit vehicle at near-side stops (1.9%) versus far-side stops (0.1%).

**Table 5: Results for data collection 1, pedestrian crossing behaviors**

| Pedestrian crossing behaviors | Far-Side | Near-Side | Far-Side | Near-Side | Statistic | df | p-value |
|---|---|---|---|---|---|---|---|
| *Dependent variable* | *Sample size* | | *Proportion (mid-block)* | | *Fisher's exact test* | | |
| Crossing location: Mid-block (vs. at an intersection or a marked crossing) | 907 | 208 | 0.0342 | 0.0240 | 0.70 | -- | 0.663 |
| *Dependent variable* | *Sample size* | | *Proportion (true)* | | *Fisher's exact test* | | |
| Pedestrian behavior: Was outside of the crosswalk markings for most if not all of the crossing | 907 | 208 | 0.0397 | 0.0433 | 1.09 | -- | 0.845 |
| Pedestrian behavior: Changed speed (e.g., walk to run, or run to walk) | 907 | 208 | 0.0232 | 0.0192 | 0.83 | -- | 1.000 |
| Pedestrian behavior: Paused in the middle of the street | 907 | 208 | 0.0077 | 0.0096 | 1.25 | -- | 0.678 |
| Pedestrian behavior: Seemed distracted by phone or something else | 907 | 208 | 0.0066 | 0.0096 | 1.46 | -- | 0.648 |
| Pedestrian behavior: Crossed just in front of the transit vehicle (if present) | 907 | 208 | 0.0011 | 0.0192 | 17.70 | -- | 0.005 |
| Pedestrian behavior: Crossed just behind the transit vehicle (if present) | 907 | 208 | 0.0099 | 0.0048 | 0.48 | -- | 0.698 |

*Pedestrian Safety: Pedestrian–Vehicle Conflicts*

Three additional pedestrian safety dependent variables related to pedestrian–vehicle conflicts were also assessed: (a) *Driver reactions* to the conflict, categorized into no obvious reaction versus some other reaction (driver fully stopped, driver slowed down, driver sped up, driver swerved); (b) *Pedestrian reactions* to the conflict, categorized into no obvious reaction versus some other reaction (stopped and waited for the vehicle, slowed down to avoid a collision, sped up or ran to avoid a collision, changed direction); and (c) *Conflict severity*, calculated as the encroachment time: the absolute value of difference between when the pedestrian and the vehicle were at the conflict point. Note that smaller values (less time) imply more severe conflicts. Results are shown in Table 6.

There were some measurable and significant differences in *driver and pedestrian reactions*. Drivers were more likely to have some other reaction when involved in conflicts at near-side stops (52%) than at far-side stops (25%). Conversely, more conflicts with some other pedestrian reaction were observed for far-side stops (62%) than for near-side stops (43%), although the difference was not statistically significant. Given the most common "other" reactions (see Table 1), it appears that drivers stopped/slowed for pedestrians more often during conflicts at near-side stops, whereas pedestrians stopped/slowed more often for drivers during conflicts at far-side stops. A minor difference in *conflict severity* was observed: conflicts were slightly more severe at far-side stops (ET of 1.81 sec) than at near-side stops (ET of 2.00 sec), but the difference was modest and not statistically significant. Overall, it appears that the small sample size of conflicts limited the ability to draw definitive conclusions, especially concerning conflict severity.

**Table 6: Results for data collection 1, pedestrian–vehicle conflicts**

| Pedestrian–vehicle conflicts | Far-Side | Near-Side | Far-Side | Near-Side | Statistic | df | p-value |
|---|---|---|---|---|---|---|---|
| *Dependent variable* | Sample size | | Proportion (other) | | Fisher's exact test | | |
| Driver reaction: Other (vs. no obvious reaction) | 36 | 23 | 0.2500 | 0.5217 | 3.20 | -- | 0.051 |
| Pedestrian reaction: Other (vs. no obvious reaction) | 34 | 23 | 0.6176 | 0.4348 | 0.48 | -- | 0.190 |
| *Dependent variable* | Sample size | | Mean (#) | | Welch's t-test | | |
| Encroachment time (sec): abs(vehicle time − pedestrian time at the conflict point) | 32 | 21 | 1.8125 | 2.0000 | -0.71 | 42 | 0.482 |

**Data Collection 2**

*Pedestrian Safety: Pedestrian–Right-Turning Vehicle Conflicts*

Four types of dependent variables related to pedestrian safety were analyzed: (a) Whether the *pedestrian crossing location* was at an intersection or away from the crosswalk (i.e., mid-block or through the middle of the intersection); (b) The *driver's reaction* to the conflict, categorized into no obvious reaction versus some other reaction (fully stopped, slowed down, sped up, swerved); (c) The *pedestrian's reaction* to the conflict, categorized into no obvious reaction versus some other reaction (stopped at waited for the vehicle, slowed down to avoid collision, sped up to avoid collision, ran to avoid collision, changed direction); and (d) *Conflict severity*, measured as encroachment time: the absolute value of difference between when the pedestrian and the vehicle were at the conflict point. Again, smaller values (less time) imply more severe conflicts. Results for all conflicts (defined in data collection 2 as ET ≤ 10 sec) are shown in Table 7. For more comparability with data collection 1, high-severity conflicts were identified using a 3 sec threshold; results are shown in Table 8.

There were noticeable differences in *pedestrian crossing locations*. More pedestrian crossings happened away from the crosswalk at near-side stops than at far-side stops. While were was a significant difference for all conflicts (5.5% versus 1.4%), the difference was of a similar magnitude (6.1% versus 2.4%) but no longer statistically significant for high-severity conflicts. In summary, it appears that pedestrians were more likely to cross away from the crosswalk at corners with near-side transit stops.

Results for *driver and pedestrian reactions* to right-turn conflicts depended somewhat on the definition of a conflict. Drivers were much more likely to have some other reaction (usually slowing down or stopping fully) for high-severity conflicts at near-side stops (82%) than for far-side stops (59%); there was no significant difference when considering all conflicts. Other pedestrian reactions (more varied, according to Table 2) were slightly more common for conflicts at far-side stops than those at near-side stops (11% versus 8% for all potential conflicts, 19% versus 15% for high severity conflicts), but the differences were not statistically significant. In summary, it appears that right-turning drivers stopped/slowed for pedestrians more often during high severity conflicts at near-side stops, whereas there was some (albeit not convincing) evidence that pedestrians were less likely to have no reaction during conflicts at far-side stops.

Similarly, the differences in *conflict severity and encroachment time* seemed to depend on the threshold for a conflict. For all potential conflicts, there was no difference in average encroachment time for near-side stops compared to far-side stops. On the other hand, when looking just at high severity conflicts, the encroachment time was significantly shorter (the conflict was more severe) for those taking place on corners with far-side stops (2.31 sec) as compared to near-side stops (2.61 sec). In summary, this suggests that high-severity conflicts tend to be more severe at far-side stops than at near-side stops.

**Table 7: Results for data collection 2, pedestrian–right-turning vehicle conflicts (max 10 sec)**

| *Pedestrian–vehicle conflicts (max 10 sec)* | Far-Side | Near-Side | Far-Side | Near-Side | Statistic | df | p-value |
|---|---|---|---|---|---|---|---|
| *Dependent variable* | *Sample size* | | *Proportion (away)* | | *Fisher's exact test* | | |
| Crossing location: Away from the crosswalk (mid-block or middle of the intersection) (vs. in the crosswalk or the crosswalk area) | 661 | 146 | 0.0136 | 0.0548 | 4.19 | -- | 0.005 |
| *Dependent variable* | *Sample size* | | *Proportion (other)* | | *Fisher's exact test* | | |
| Driver reaction: Other (vs. no obvious reaction) | 661 | 146 | 0.4932 | 0.4521 | 0.85 | -- | 0.410 |
| Pedestrian reaction: Other (vs. no obvious reaction) | 661 | 146 | 0.1074 | 0.0753 | 0.68 | -- | 0.291 |
| *Dependent variable* | *Sample size* | | *Proportion (true)* | | *Fisher's exact test* | | |
| Conflict severity: High (0-3 sec) | 661 | 146 | 0.1861 | 0.2260 | 1.28 | -- | 0.297 |
| Conflict severity: Mild (4-5 sec) | 661 | 146 | 0.2935 | 0.3288 | 1.18 | -- | 0.425 |
| Conflict severity: Low (6-10 sec) | 661 | 146 | 0.5204 | 0.4452 | 0.74 | -- | 0.120 |
| *Dependent variable* | *Sample size* | | *Mean (#)* | | *Welch's t-test* | | |
| Encroachment time (sec): abs(vehicle time − pedestrian time at the conflict point) | 661 | 146 | 5.5900 | 5.5616 | 0.13 | 207 | 0.896 |

**Table 8: Results for data collection 2, pedestrian–right-turning vehicle conflicts (max 3 sec)**

| *Pedestrian–vehicle conflicts (max 3 sec)* | Far-Side | Near-Side | Far-Side | Near-Side | Statistic | df | p-value |
|---|---|---|---|---|---|---|---|
| *Dependent variable* | *Sample size* | | *Proportion (away)* | | *Fisher's exact test* | | |
| Crossing location: Away from the crosswalk (mid-block or middle of the intersection) (vs. in the crosswalk or the crosswalk area) | 123 | 33 | 0.0244 | 0.0606 | 2.56 | -- | 0.285 |
| *Dependent variable* | *Sample size* | | *Proportion (other)* | | *Fisher's exact test* | | |
| Driver reaction: Other (vs. no obvious reaction) | 123 | 33 | 0.5935 | 0.8182 | 3.06 | -- | 0.024 |
| Pedestrian reaction: Other (vs. no obvious reaction) | 123 | 33 | 0.1870 | 0.1515 | 0.78 | -- | 0.800 |
| *Dependent variable* | *Sample size* | | *Mean (#)* | | *Welch's t-test* | | |
| Encroachment time (sec): abs(vehicle time − pedestrian time at the conflict point) | 123 | 33 | 2.3089 | 2.6061 | -2.42 | 76 | 0.018 |

## Data Collection 3

### *Pedestrian Safety: Pedestrian Crossing Behaviors*

Three types of dependent variables related to pedestrian crossing behaviors were investigated: (a) Whether the *pedestrian crossing location* was at an intersection or away from the crosswalk (i.e., mid-block or through the middle of the intersection); (b) Whether the pedestrian stayed within the *crosswalk markings* for most or all of the crossing, whether they were outside of the crosswalk markings for most or all of the crossing, or whether they were both inside and outside for part of the crossing; and (c) Multiple observations related to *pedestrian crossing behaviors and obstacles*, such as: changed speed, paused in the middle of the street, seemed distracted, or encountered a car blocking the crosswalk. Results are shown in Table 9.

There were statistically significant differences in *pedestrian crossing locations*: Many more pedestrian crossings happened away from the crosswalk at locations with far-side stops (3.8%) than those with near-side stops (0.7%). In summary, it appears that pedestrians were more likely to cross away from the crosswalk when crossing to/from a corner with a far-side transit stop.

Most *pedestrian crossing behaviors* and situations were not significantly more or less common for near-side versus far-side transit stops: most adherence to crosswalk markings, changing speed, appearing distracted, and car blocking the crosswalk. However, more pedestrians crossed while mostly outside of the crosswalk markings at near-side stops (8.4%) vs. far-side stops (5.9%). Also, it was somewhat more common to observe pedestrians pausing in the middle of the street for crossings linking to corners with far-side stops (2.5%) than for near-side stops (0.9%). In summary, crossings with near-side stops were more likely to observe pedestrians crossing mostly outside of the crosswalk markings, while crossings with far-side stops were more likely to observe pedestrians pausing in the middle of the street.

**Table 9: Results for data collection 3, pedestrian crossing behaviors**

| **Pedestrian crossing behaviors** | Far-Side | Near-Side | Far-Side | Near-Side | Statistic | df | p-value |
|---|---|---|---|---|---|---|---|
| *Dependent variable* | Sample size | | Proportion (mid-block) | | Fisher's exact test | | |
| Crossing location: Away from the crosswalk (mid-block or middle of the intersection) (vs. in the crosswalk or the crosswalk area) | 2,133 | 439 | 0.0380 | 0.0068 | 0.17 | -- | 0.000 |
| *Dependent variable* | Sample size | | Proportion (true) | | Fisher's exact test | | |
| Crosswalk markings: Stayed within the crosswalk markings for all or almost the whole crossing | 2,172 | 439 | 0.8476 | 0.8542 | 1.05 | -- | 0.770 |
| Crosswalk markings: Stepped outside of the crosswalk markings for part of the crossing | 2,172 | 439 | 0.0829 | 0.0911 | 1.11 | -- | 0.572 |
| Crosswalk markings: Was outside of the crosswalk markings for most if not all of the crossing | 2,172 | 439 | 0.0589 | 0.0843 | 1.47 | -- | 0.053 |
| Pedestrian behavior: Changed speed (e.g., walk to run, or run to walk) | 2,172 | 439 | 0.0557 | 0.0729 | 1.33 | -- | 0.180 |
| Pedestrian behavior: Paused in the middle of the street | 2,172 | 439 | 0.0253 | 0.0091 | 0.35 | -- | 0.034 |
| Pedestrian behavior: Seemed distracted by phone or something else | 2,172 | 439 | 0.0041 | 0.0023 | 0.55 | -- | 1.000 |
| Crossing obstacle: Car blocking the crosswalk | 2,172 | 439 | 0.0479 | 0.0387 | 0.80 | -- | 0.457 |

**DISCUSSION**

Recall the research objective to assess the impacts of near-side versus far-side transit stop location on both traffic operations and pedestrian safety at signalized intersections. Analysis of three different datasets—transit vehicle stop events and transit passenger behaviors (data collection 1), pedestrians involved in conflicts with right-turning vehicles (data collection 2), and pedestrian street crossing behaviors (data collection 3)—helped to identify key findings and policy implications, as discussed in the following subsections.

**Key Findings**

*Traffic operations*: For public transit operations, arrival delays and impacts were more likely and more severe at near-side stops, but departure delays were much more likely and impactful at far-side stops. For other traffic, transit vehicles delayed other vehicles much more often and more significantly at near-side stops, whereas somewhat more passing was observed at far-side stops. These results make sense and match findings from the literature and expectations from transit agencies. Reducing delays to other traffic is likely a major reason why many more transit stops in Utah are placed on the far-side of intersections. However, there is a tradeoff with transit operations: transit vehicles are often delayed (17% of the time) when trying to leave the stop and rejoin traffic.

*Pedestrian safety, crossing location and behaviors*: Results are somewhat inconclusive. Transit riders (data collection 1) and pedestrians crossing the street (data collection 3) were more likely to cross mid-block or away from the crosswalk in the presence of far-side transit stops than near-side transit stops; however, more pedestrians involved in a right-turn conflict (data collection 2) were crossing away from the crosswalk at corners with near-side transit stops (but, consider the smaller sample size). Perhaps far-side stops encourage more mid-block crossings because they are placed further from the intersection than near-side stops, or pedestrians do not want to backtrack to the intersection (after having ridden past it) to cross. Few pedestrian behaviors while crossing showed significant differences. Transit riders crossing in front of the transit vehicle (data collection 1), and pedestrians crossing mostly outside of the crosswalk markings (data collection 3), were both observed more often at near-side stops than at far-side stops. Near-side stops may be located close to the stop bar, and transit stop events may be more likely to coincide with a red indication for the transit vehicle. Thus, pedestrians may be crossing legally in the crosswalk (or cutting the corner and walking near but outside the crosswalk markings) while the walk indication is on for crossing that street. More detailed observations of pedestrian trajectories and traffic signal timing would be necessary to substantiate these potential explanations.

*Pedestrian safety, pedestrian–vehicle conflicts*: Despite the small sample sizes, all signs point towards potential pedestrian safety concerns associated with far-side transit stops. Drivers were less likely to stop or slow during conflicts with pedestrians at far-side stops. Although not significant, pedestrians were slightly more likely to have to react (take evasive action) in some way at far-side stops. When considering encroachment time (as an inverse measure of conflict severity), conflicts or near-misses were slightly more severe (less time between road users) at far-side stops. A possible explanation is that, since transit lines usually run along the main street of an intersection, vehicles may be traveling faster at (or turning right more quickly onto) the far-side leg of the main street (where pedestrians are crossing on their way to/from a far-side transit stop), but traveling slower when approaching (or turning right off of) the near-side leg of the main street (where pedestrians are crossing on their way to/from a near-side transit stop). Again, this hypothesis would need to be corroborated through more detailed observations of pedestrian and vehicle speeds, trajectories, interactions.

**Policy Implications**

What do these findings recommend regarding transit stop placement to improve pedestrian safety and traffic operations at signalized intersections? Unfortunately, the implications differ and there are tradeoffs for near-side and far-side transit stops. Regarding *traffic operations*, far-side stops are better for general traffic operations, because fewer other vehicles are impacted by transit stop events. For transit operations, near-side stops delay transit vehicles when approaching the stop, but far-side stops delay transit vehicles when departing the stop. Which is worse depends upon situational factors (*24*). Given the predominance and preference for far-side transit stops (at least in Utah), departure delays could be reduced by implementing "yield-to-transit" laws (*29*) and/or constructing curb extensions to house in-lane far-side transit stops (*24*).

Regarding *pedestrian safety*, despite some contradictory or inconclusive findings regarding pedestrian crossing locations and behaviors, overall the evidence pointed towards far-side transit stops being worse for pedestrian safety. Specifically, mid-block crossings were more common, drivers were less likely to stop/slow for pedestrians, and (especially) conflicts with vehicles were more severe (less time between road users) at far-side stops. These results suggest that adverse safety outcomes at far-side transit stops may be affected by both pedestrian and driver behaviors and actions. These observational findings corroborate evidence from Utah-based crash analyses (*14–16*) about the adverse impacts of far-side transit stops on pedestrian crashes.

What should be done, if far-side stops are better (or can be made better) for traffic/transit operations, but far-side stops are also worse for pedestrian safety? First, strategies could try to make far-side transit stops safer for pedestrians: curb extensions and tighter corner radii, leading pedestrian intervals and prohibiting right turns on red, and intersection traffic calming. However, these may negatively impact operations. Second, pedestrian safety could be prioritized over traffic operations, and near-side transit stops could be recommended in certain situations. Many states (including Utah) have Vision Zero plans and policies, which suggests that safety should be a higher priority than operations in many situations (potentially including transit stop placement). A combination of both approaches—improving safety at far-side stops and accepting some operational inefficiencies at near-side stops—might be the best solution.

**Limitations and Future Work**

A variety of study limitations could be addressed through future work. The limited geographic scope (Utah), small sample size for some analyses, and varied methods of data collection somewhat negatively impacted the statistical power to detect significant differences and the generalizability of the findings. The simple bivariate analyses did not account for various potentially influential factors such as time-of-day, weather, land use, and intersection characteristics. Future research should extend to more locations and longer timeframes, incorporate controls for these variables, and examine the distance of transit stops from intersections to better understand their impact on pedestrian behavior. Additionally, employing techniques like computer vision to analyze the trajectories and speeds of road users, and using microsimulations for traffic and transit operations, could provide deeper insights into how transit stop placement affects pedestrian safety and traffic efficiency.

**ACKNOWLEDGEMENTS**

Special thanks to several undergraduate students at Utah State University—Brian Beard, Sadie Boyer, William Bouck, Melissa Brown, Erica Drollinger, Alyssa Gaither, Preston Goodrich, Max Haehnel, Matthew Hall, Emilie Hill, Chandler Hokanson, Bradley Howell, Bailey Nielson, Aleks Paskett, Thomas Shaw, Jordan Taft, Kyle Wariner, Peyton Webb, and Seth Wilcox—for helping to collect some of the data used in this project.

The work presented in this paper was conducted with support from Utah State University, the Utah Transit Authority, the Utah Department of Transportation, and the Mountain-Plains Consortium, a University Transportation Center funded by the U.S. Department of Transportation. The contents of this report reflect the views of the authors, who are responsible for the facts and the accuracy of the information presented. The contents do not necessarily reflect the views, opinions, endorsements, or policies of the US Department of Transportation, the Utah Department of Transportation, or the Utah Transit Authority.

**AUTHOR CONTRIBUTION STATEMENT**

The authors confirm contribution to the paper as follows: study conception and design: PAS, MM, FS, AS; data collection: FS, AS, PS, MM; analysis and interpretation of results: PAS, FS, AS, MM; draft manuscript preparation: FS, AS, PS. All authors reviewed the results and approved the final version of the manuscript.

27. Singleton, P. A., Mekker, M., Gaither, A., Subedi, A., & Islam, A. (2023b). *Right-turn safety for walking/bicycling: Impacts of curb/corner radii and other factors* (No. UT-23.09). Utah Department of Transportation. https://rosap.ntl.bts.gov/view/dot/72595
28. US Census Bureau. (2021). *Change in resident population of the 50 states, the District of Columbia, and Puerto Rico: 1910 to 2020*. https://www2.census.gov/programs-surveys/decennial/2020/data/apportionment/population-change-data-table.pdf
29. Oregon Revised Statutes (ORS). (2024). Failure to yield right of way to transit bus (§ 811.167). https://oregon.public.law/statutes/ors_811.167
20